\documentstyle[prl,aps]{revtex}
\begin{document}
\wideabs{
\title{Antiproton Production in 11.5 A GeV/c Au+Pb Nucleus-Nucleus Collisions}
\maketitle
% \draft command makes pacs numbers print
\draft
% repeat the \author\address pair as needed
\begin{center}
\bigskip
\mbox{T.A. Armstrong                \unskip,$^{7}$}
\mbox{K.N. Barish                   \unskip,$^{11,\ast}$}
\mbox{M.J. Bennett                  \unskip,$^{11,\dag}$}
\mbox{S.J. Bennett                  \unskip,$^{10}$}
\mbox{A. Chikanian                  \unskip,$^{11}$}
\mbox{S.D. Coe                      \unskip,$^{11}$}
\mbox{T.M. Cormier                  \unskip,$^{10}$}
\mbox{R. Davies                     \unskip,$^{8}$}
\mbox{G.De Cataldo                  \unskip,$^{1}$}
\mbox{P. Dee                        \unskip,$^{10,\ddag}$}
\mbox{G.E. Diebold                  \unskip,$^{11}$}
\mbox{C.B. Dover                    \unskip,$^{2,\S}$}
\mbox{P. Fachini                    \unskip,$^{10}$}
\mbox{L.E. Finch                    \unskip,$^{11}$}
\mbox{N.K. George                   \unskip,$^{11}$}
\mbox{N. Giglietto                  \unskip,$^{1}$}
\mbox{S.V. Greene                   \unskip,$^{9}$}
\mbox{P. Haridas                    \unskip,$^{6}$}
\mbox{J.C. Hill                     \unskip,$^{4}$}
\mbox{A.S. Hirsch                   \unskip,$^{8}$}
\mbox{H.Z. Huang                    \unskip,$^{3,\|}$}
\mbox{B. Kim                        \unskip,$^{10}$}
\mbox{B.S. Kumar                    \unskip,$^{11,\P}$}
\mbox{J.G. Lajoie                   \unskip,$^{11}$}
\mbox{R.A. Lewis                    \unskip,$^{7}$}
\mbox{Q. Li                         \unskip,$^{10}$}
\mbox{Y. Li                         \unskip,$^{10}$}
\mbox{B. Libby                      \unskip,$^{4,\ast\ast}$}
\mbox{R.D. Majka                    \unskip,$^{11}$}
\mbox{M.G. Munhoz                   \unskip,$^{10}$}
\mbox{J.L. Nagle                    \unskip,$^{11,\dag\dag}$}
\mbox{I.A. Pless                    \unskip,$^{6}$}
\mbox{J.K. Pope                     \unskip,$^{11}$}
\mbox{N.T. Porile                   \unskip,$^{8}$}
\mbox{C.A. Pruneau                  \unskip,$^{10}$}
\mbox{M.S.Z. Rabin                  \unskip,$^{5}$}
\mbox{A. Raino                      \unskip,$^{1}$}
\mbox{J.D. Reid                     \unskip,$^{7,\ddag\ddag}$}
\mbox{A. Rimai                      \unskip,$^{8,\S\S}$}
\mbox{F.S. Rotondo                  \unskip,$^{11}$}
\mbox{J. Sandweiss                  \unskip,$^{11}$}
\mbox{R.P. Scharenberg              \unskip,$^{8}$}
\mbox{A.J. Slaughter                \unskip,$^{11}$}
\mbox{G.A. Smith                    \unskip,$^{7}$}
\mbox{P. Spinelli                   \unskip,$^{1}$}
\mbox{B.K. Srivastava               \unskip,$^{8}$}
\mbox{M.L. Tincknell                \unskip,$^{8}$}
\mbox{W.S. Toothacker               \unskip,$^{7}$}
\mbox{G. Van Buren                  \unskip,$^{6}$}
\mbox{W.K. Wilson                   \unskip,$^{10}$}
\mbox{F.K. Wohn                     \unskip,$^{4}$}
\mbox{E.J. Wolin                    \unskip,$^{11,\|\:\|}$}
\mbox{K. Zhao                       \unskip$^{10}$}

\vskip \baselineskip
\centerline{(The E864 Collaboration)}
\it
  $^{1}$University of Bari/INFN, Bari, Italy \break
  $^{2}$Brookhaven National Laboratory, Upton, New York 11973 \break
  $^{3}$University of California at Los Angeles, Los Angeles, 
	  California 90024 \break 
  $^{4}$Iowa State University, Ames, Iowa 50011 \break
  $^{5}$University of Massachusetts, Amherst, Massachusetts 01003 \break
  $^{6}$Massachusetts Institute of Technology, Cambridge, 
	  Massachusetts 02139 \break
  $^{7}$Pennsylvania State University, University Park, 
	  Pennsylvania 16802 \break
  $^{8}$Purdue University, West Lafayette, Indiana 47907 \break
  $^{9}$Vanderbilt University, Nashville, Tennessee 37235 \break
  $^{10}$Wayne State University, Detroit, Michigan 48201 \break
  $^{11}$Yale University, New Haven, Connecticut 06520 \break

\end{center}

\begin{abstract}
% insert abstract here
We present the first results from the E864 collaboration on the production 
of antiprotons in 10\% central 11.5 A GeV/c Au+Pb nucleus collisions at the 
Brookhaven AGS.  
We report invariant multiplicities for antiproton production in the kinematic 
region 
$1.4<y<2.2$ and $50<p_{T}<300$ MeV/c,
and compare our data with a first collision scaling model and previously 
published results from the  
E878 collaboration. The differences between the E864 and
E878 antiproton measurements and the implications for antihyperon production 
are discussed.
\end{abstract}
% insert suggested PACS numbers in braces on next line
}

% body of paper here

The yield of antiprotons ($\overline{p}$) in high energy heavy-ion collisions 
is of
considerable interest for several reasons. Models of heavy-ion collisions
that include a quark-gluon plasma (QGP) phase predict that the production of
antimatter will be enhanced in these collisions due to the lower 
quark-antiquark
production threshold relative to that of a baryon-antibaryon pair 
\cite{qgprefs}.
Enhanced production of antimatter may also indicate strong, density 
dependent mean field effects \cite{mean}.  The observed yield is, 
however, a result of both production and subsequent annihilation. 
Detailed study of $\overline{p}$ production has been
proposed as an indirect way of measuring the baryon density in these 
collisions \cite{density}.
Finally, $\overline{p}$ measurements at the AGS may contain a large feed-down
contribution from the decay of the antilambda, $\overline{\Lambda} 
\rightarrow \overline{p}+\pi^{+}$, 
as well as other antihyperons ($\overline{Y}$).  
By comparing results from experiments with different
sensitivities to $\overline{p}$'s from these decays, we may be able to 
infer the 
relative production of antihyperons and $\overline{p}$'s in Au+Pb collisions.
                        
Experiment 864 is a high rate, large acceptance spectrometer designed to 
search for novel 
forms of matter created in heavy ion collisions.
The spectrometer consists of two dipole bending magnets (M1 and M2), 
with time-of-flight
(TOF) hodoscopes and straw tube tracking chambers downstream of the 
second magnet
(see Figure \ref{spect}).
The scintillation counter hodoscopes (H1, H2 and H3) provide space points 
for tracking, as well as redundant charge and TOF 
measurements. The TOF resolution is $\sim$130 ps in each plane.  
The straw tube chambers (S2 and S3) provide high precision space points. 
The mass resolution
of the spectrometer is between 3 and 5\% in the kinematic region explored 
in this data set. 
At the end of the apparatus is a lead/scintillating fiber hadronic calorimeter, 
which is used to confirm the energy of the particle determined by the 
tracking detectors \cite{Claude}.
Finally, a straw tube chamber (S1) located between the spectrometer magnets  
provides additional background rejection. 
The uninteracted beam is contained above the experiment
in a large vacuum chamber.  The centrality (impact parameter) of 
the collision is determined by a segmented scintillation
counter located near the target, which measures charged particle multiplicity 
in the polar angular range of 16.6$^{o}$
to 45$^{o}$ with respect to the incident beam.  For this analysis 
we selected events with the 10\% largest
pulse heights in the multiplicity array, roughly corresponding to 
the 10\% most central events or an impact parameter 
%$b\stackrel{ \scriptstyle < }{\sim}4.7$ fm. 
$b \leq 4.7$ fm.
A complete description of the apparatus is in preparation \cite{e864nim}.

The data presented in this paper are derived from 20.1 million 10\% 
central Au+Pb interactions collected during
the 1994 run with 5\%, 10\% and 20\% Pb targets.  
For the 1994 run, the experimental apparatus was not complete.  
The calorimeter
was $\frac{1}{4}$ complete, two layers of the S3 straw array were 
only $\frac{1}{3}$
complete, and S1 was not in place.  The calorimeter was stacked to have optimal 
acceptance for neutral particles, and thus was not used in this analysis. 
An average of five tracks per event was found in the spectrometer.  The mass
of a particle is calculated as $m = Z \frac{r}{\beta\gamma}$
where the rigidity $r$ is reconstructed from the downstream fit 
of the track in the bend plane, 
and the charge $Z$ and velocity $\beta$ are measured by the hodoscopes.  
Antiproton candidates are selected by a set of quality cuts on the fits 
to the particle track in the detector.  
Additional cuts are made to exclude particles whose
back-projection intercepts the collimator in the first spectrometer magnet 
and to demand a charge measurement consistent 
with $Z=1$ in each hodoscope.  The resulting mass distribution is fit to 
a combination
of a Gaussian signal and linear background in the mass region of the 
$\overline{p}$. 
The ratio of the number of signal counts in $\pm3\sigma$ about the fixed 
$\overline{p}$ peak 
to the number of background counts is $\sim3$.  
The measured $\overline{p}$ yields are corrected for the experimental 
acceptance and efficiencies in each
rapidity and $p_{T}$ bin.  Since the TOF information from all three 
hodoscopes is required for a track
in the spectrometer, an occupancy dependent correction is made to 
account for the fact that if a faster
track hits a hodoscope slat first, later tracks could be lost, 
because only the first time will 
be recorded.
                                              
The $\overline{p}$ invariant multiplicities measured in E864 are shown in 
Figure \ref{pb_invmlt}.  
The measured multiplicities are approximately flat over the $p_{T}$ range 
where the experiment has acceptance, and 
correspond to a level of $1.5 \times 10^{-2}$ GeV$^{-2}$c$^{2}$ at 
midrapidity ($y=1.6$).
Invariant multiplicities for $\overline{p}$ production as measured by the E878 
collaboration \cite{e878_prl} are also shown in Figure \ref{pb_invmlt}. 
The E878 data have been scaled up by a factor of 1.5 to account for the 
lower beam momentum of 10.8 A GeV/c using the procedure in \cite{arc_pap}. 
We estimate that there is a 15\% systematic uncertainty in
this energy scaling based on a comparison with fits
to higher energy pp data \cite{costales}.  It should also be noted that the E878 measurements 
are for Au+Au nucleus collisions;
however, the difference between the Au (Z=79, A=197) and Pb (Z=82, A=208) 
target nuclei is negligible. 
We estimate the systematic errors in our measurements to be 20\%, dominated by
our understanding of the experimental acceptance and track quality cut 
efficiencies. 
E878 reports a systematic error of 30\% on their measurements \cite{e878_prl}.  
Figure \ref{pb_invmlt} shows that the E864 measurements are consistently higher
than their E878 counterparts.  

Since the E864 data are approximately flat as a function of $p_{T}$ in 
each rapidity interval, they 
are averaged in each rapidity range and extrapolated to 
$p_{T}=0$.  
If we assume a Boltzmann shape to the $\overline{p}$ distribution with 
a temperature parameter
of 200 MeV (similar to preliminary measurements for $\overline{p}$'s 
by the E866 collaboration  
\cite{e866_pbar}), this extrapolation could underestimate the invariant 
multiplicity at $p_{T}=0$ by 6\%.
It should be noted that we cannot rule out a drastic change in $\overline{p}$
production between $0<p_{T}<50$ MeV/c.
If there is such a low-$p_{T}$ dependence to $\overline{p}$ production, 
this extrapolation will not be valid.
Figure \ref{rapid_width} shows the E864 extrapolations to $p_{T}=0$ along 
with the scaled E878
measurements.  The E878 measurements yield a 
rapidity width of $\sigma_{y}=0.62\pm0.03$ \cite{e878_prl}, 
while the E864 data yield a width of $0.49\pm.05$.
Using the E864 measurement extrapolated to $p_{T}=0$, we can quantify the 
difference between E864 and E878.
At midrapidity the ratio of invariant multiplicities is 
$3.96\pm 0.42 \mathrm{(stat.)} ^{+5.08} _{-1.77} \mathrm{(syst.)}$.
This ratio decreases as one moves away from central rapidity.
%Using Gaussian fits to the data, we find that the 
%the ratio of integrated yields at $p_{T}=0$ is $3.37\pm 0.51 
%\mathrm{(stat.)} ^{+4.34} _{-1.52} \mathrm{(syst.)}$.
%We discuss this ratio further in a later section.
We will proceed by comparing the measured $\overline{p}$ production in 
E864 with a simple model, 
and then return to the discrepancy between the E864 and E878 measurements and 
its implications for antihyperon production.   

If we assume a model in which $\overline{p}$'s 
are produced only in first collisions between target and projectile 
nucleons, and are not annihilated,
we can estimate the $\overline{p}$ yield 
by scaling $\overline{p}$ production in nucleon-nucleon collisions by 
the number of first collisions
in a nucleus-nucleus interaction.  A first collisions model of this 
type provides a reference level
of production for comparison, and may provide an indication as to 
whether $\overline{p}$ production
is substantially enhanced in nucleus-nucleus collisions, or suppressed 
(by annihilation). 
It has been estimated that there are typically 47 first collisions 
between nucleons in 
10\% central Au+Au(Pb) collisions \cite{shiva}.  
We estimate the $\overline{p}$ production in pp collisions using 
RQMD (v2.2) \cite{RQMD}, which is tuned to 
measured $\overline{p}$ production at higher energies and includes 
energy scaling of the cross section.  
The result of multiplying the RQMD pp invariant $\overline{p}$ 
yield at $p_{T}=0$ by the expected number of first collisions 
in a Au+Pb collision is shown in Figure \ref{rapid_width}.  
%The rapidity width of the first collisions model is $.42$, somewhat 
%smaller than measured in E864.   
%In addition, the overall level of production in the first collisions 
%model is higher than measured in E864 at $p_{T}=0$. 
In pp collisions $\overline{p}$ production is peaked at low $p_{T}$, and 
the RQMD distribution can be described by a Boltzmann temperature of $\sim$80 MeV. 
In nucleus-nucleus collisions, rescattering
and annihilation are expected to broaden the distribution in rapidity 
and $p_{T}$.  In Figure
\ref{rapid_width} we also show the invariant yield at $p_{T}=0$ assuming 
the same integrated yield 
(47 times the RQMD pp level) and a Boltzmann temperature of 200 MeV.  
In this case, the yield at $p_{T}=0$ is
somewhat lower than the yield measured in E864.  This could be an 
indication of enhanced $\overline{p}$ production 
in Au+Pb nucleus collisions that more than offsets any losses due to 
annihilation.   

In general, the $\overline{p}$'s detected could also be the decay products of   
antihyperons, such as the $\overline{\Lambda}$, $\overline{\Sigma^{0}}$, 
and the $\overline{\Sigma^{+}}$.  
The decay of the $\overline{\Sigma^{0}}$ will produce additional 
$\overline{\Lambda}$'s which will be 
indistinguishable from those created in the primary collision.  
The decay of the $\overline{\Lambda}$ and the $\overline{\Sigma^{+}}$ will 
produce $\overline{p}$'s whose production vertices do not coincide with the 
location of the primary interaction between
the two nuclei.  Therefore, the degree to which $\overline{p}$'s 
from these decays contribute to a measurement of
$\overline{p}$ production will vary among experiments.  

Due to its large acceptance, the E864 spectrometer will detect 
$\overline{p}$'s from $\overline{Y}$ decay.
E864 does not have sufficient vertical resolution
to reject $\overline{p}$'s from $\overline{Y}$ decay based on 
the vertical projection
of a particle to the target, and the analysis cuts do not preferentially reject 
antiprotons from $\overline{Y}$ decay.
Therefore, the $\overline{p}$'s detected in E864 are a combination 
of primary $\overline{p}$'s and 
$\overline{p}$'s from $\overline{Y}$ decay, in a ratio that 
reflects their production ratio.  
The E878 collaboration have also 
evaluated the acceptance of their spectrometer for 
feed-down from $\overline{Y}$ decay. 
At midrapidity the acceptance for $\overline{p}$'s 
from $\overline{\Lambda}$ and $\overline{\Sigma^{0}}$ decay is  
14\% of the spectrometer acceptance for primordial $\overline{p}$'s,
and 10\% of the $\overline{p}$ 
acceptance for $\overline{\Sigma^{+}}$ decays. 

Since both E878 and E864 measure a different combination
of primordial $\overline{p}$ production and feed-down from
$\overline{Y}$ decay, we can in principle separate the
two components if we make two explicit assumptions:
both E864 and E878 understand their systematic errors, and the
entire difference between the two experiments can be 
attributed to antihyperon feed-down.  
It is important to note that in energy scaling the E878 results
we have implicitly assumed that the $\overline{Y}$'s scale
with energy by the same factor as the $\overline{p}$'s.    
A detailed statistical analysis 
of the $\overline{Y}/\overline{p}$  
ratio as a function of the E864 and E878 measurements 
(see Figure \ref{probab} for details), and the 
relevant statistical and systematic errors involved, shows that   
\begin{equation}
\left(\:\frac{\overline{Y}}{\overline{p}}\:\right) 
_{\stackrel{\scriptstyle y=1.6 }{p_{T}=0}} \approx
\left(\:\frac{\overline{\Lambda}+\overline{\Sigma^{0}}
+1.1\overline{\Sigma^{+}}}{\overline{p}}\:\right)
> \textrm{2.8  (98\% C.L.)} 
\end{equation}
while the most probable value of this ratio is $\sim5$.
The factor of 1.1 multiplying the $\overline{\Sigma^{+}}$
arises due to the different branching ratio and 
acceptance for the $\overline{\Sigma^{+}}$ compared 
to the $\overline{\Lambda}$.
This indicates an $\overline{Y}/\overline{p}$ ratio in
Au+Pb collisions that is significantly greater than one at midrapidity 
and $p_{T}=0$.  It should be noted that if the
$\overline{Y}$'s and the $\overline{p}$ are produced with different 
distributions in $y$ and $p_{T}$, then the ratio of integrated yields 
of these particles will differ from the ratio at central rapidity and $p_{T}=0$.
Preliminary results from Si+Au collisions based on 
direct measurements of $\overline{p}$ and $\overline{\Lambda}$ production 
by the E859 collaboration
also indicate a ratio of integrated yields greater than one \cite{Yeudong_Wu}.
In contrast, the ratio in pp collisions at similar energies 
is $\sim 0.2$ \cite{pp_ref}. 

An enhancement of antihyperons arises naturally
in models that include a QGP, and therefore enhanced antimatter and strangeness
production \cite{qgprefs}.  Thermal models that use a temperature and 
baryon chemical potential derived from measured particle spectra also indicate 
that the primordial $\overline{\Lambda}/\overline{p}$ ratio could be larger 
than one \cite{hgas_refs}. However, these models
are typically used to compare integrated yields while we have only inferred the 
$\overline{Y}/\overline{p}$ ratio at a point in phase space. 
  
In comparing the results of two experiments the potential exists for
differences in the overall normalization.  We
note that preliminary measurements of protons, $K^{-}$, deuterons, and $He^{3}$
in E864 are consistent with preliminary E878 measurements within the quoted 
statistical and systematic errors of both experiments \cite{Kyle_hipags}.  

In summary, E864 has measured $\overline{p}$ production about midrapidity in
Au+Pb collisions at 11.5 A GeV/c.  
The measured $\overline{p}$ yields at midrapidity and $p_{T}=0$ are larger 
than those measured by the E878 collaboration by a factor $3.96\pm 0.42 
\mathrm{(stat.)} ^{+5.08} _{-1.77} \mathrm{(syst.)}$. 
If we interpret the difference between E864 and E878 as a measure of
$\overline{Y}$ production, we can infer that 
$\overline{Y}/\overline{p}$ is much greater than one at 
midrapidity and $p_{T}=0$.
%While such an interpretation of the difference between the E864 and E878 
%results is intriguing, it is best 
%viewed as an argument that the production of antihyperons in heavy ion 
%collisions deserves further study. 

We would like to acknowledge the efforts of the AGS and Tandem 
staff in providing the beam.
This work was supported by grants from the Department of Energy (DOE) 
High Energy Physics Division, 
DOE Nuclear Division, and the National Science Foundation.  
We would like to thank Heinz Sorge for 
his assistance with the RQMD calculations and Sid Kahana for 
illuminating discussions.

% now the references. delete or change fake bibitem. delete next three
%   lines and directly read in your .bbl file if you use bibtex.

\begin{figure}
\caption{The E864 spectrometer (1994 configuration). Note that the 
calorimeter and S3 U,V layers
are incomplete. The Au beam is incident on a Pb target from the left, 
and the scale shown is in meters.
The vacuum chamber downstream of M2 is not shown in the plan view. 
See text for details.}
\label{spect}
\end{figure}

\begin{figure}
\caption{$\overline{p}$ invariant multiplicities as measured in E864 and E878,
 for 10\% central Au+Pb and Au+Au collisions. 
The E878 data have been scaled up by a factor of 1.5 to account for the lower 
beam momentum of 10.8 A GeV/c.
The errors bars are statistical only.
The inset shows the mass distribution measured in E864 in the $\overline{p}$ mass
region.  Note the logarithmic scale. } 
\label{pb_invmlt}
\end{figure}

\begin{figure}
\caption{$\overline{p}$ invariant multiplicities as extrapolated to $p_{T}=0$ 
in E864 
and measured in E878 (scaled to 11.5 A GeV/c), for 10\% central Au+Pb and 
Au+Au collisions. Fits to the
E864 and E878 data are also shown. The errors are statistical only.
Two predictions based on first collision scaling (without annihilation) are 
also indicated.}
\label{rapid_width}
\end{figure}

\begin{figure}
\caption{The probability distribution for the ratio 
$(\overline{\Lambda}+\overline{\Sigma^{0}}+1.1\overline{\Sigma^{+}})/\overline{p}$ 
at midrapidity and $p_{T}=0$ extracted from the E878 and E864 measurements.  
This distribution is generated by varying the E878 and E864 measurements within
their systematic and statistical errors. Statitiscal errors are treated as Gaussian,
while systematic errors are treated as indicating a flat range within which the 
measurements may vary.}
\label{probab}
\end{figure}

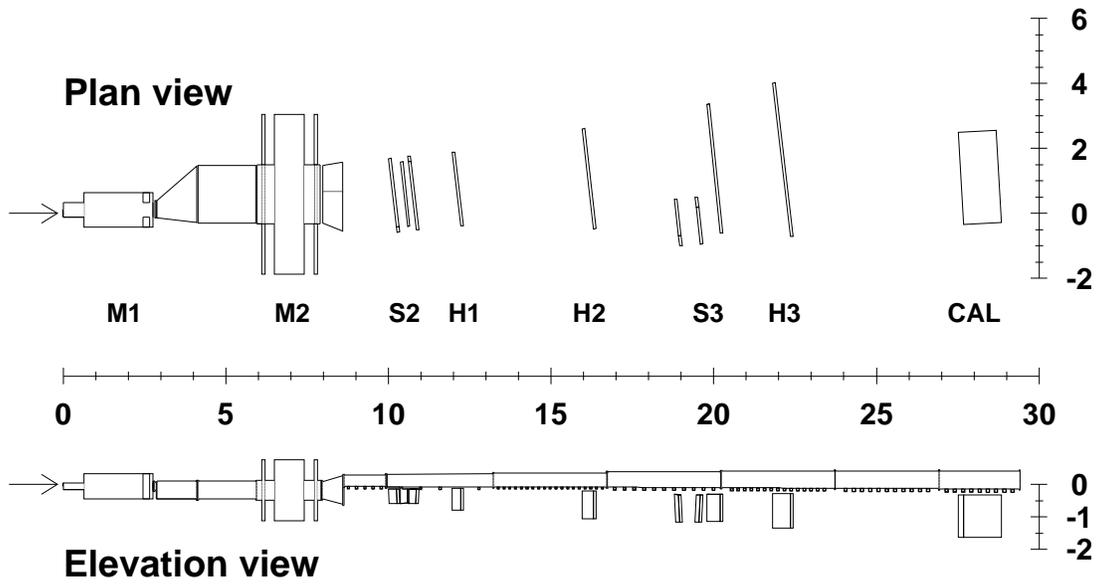
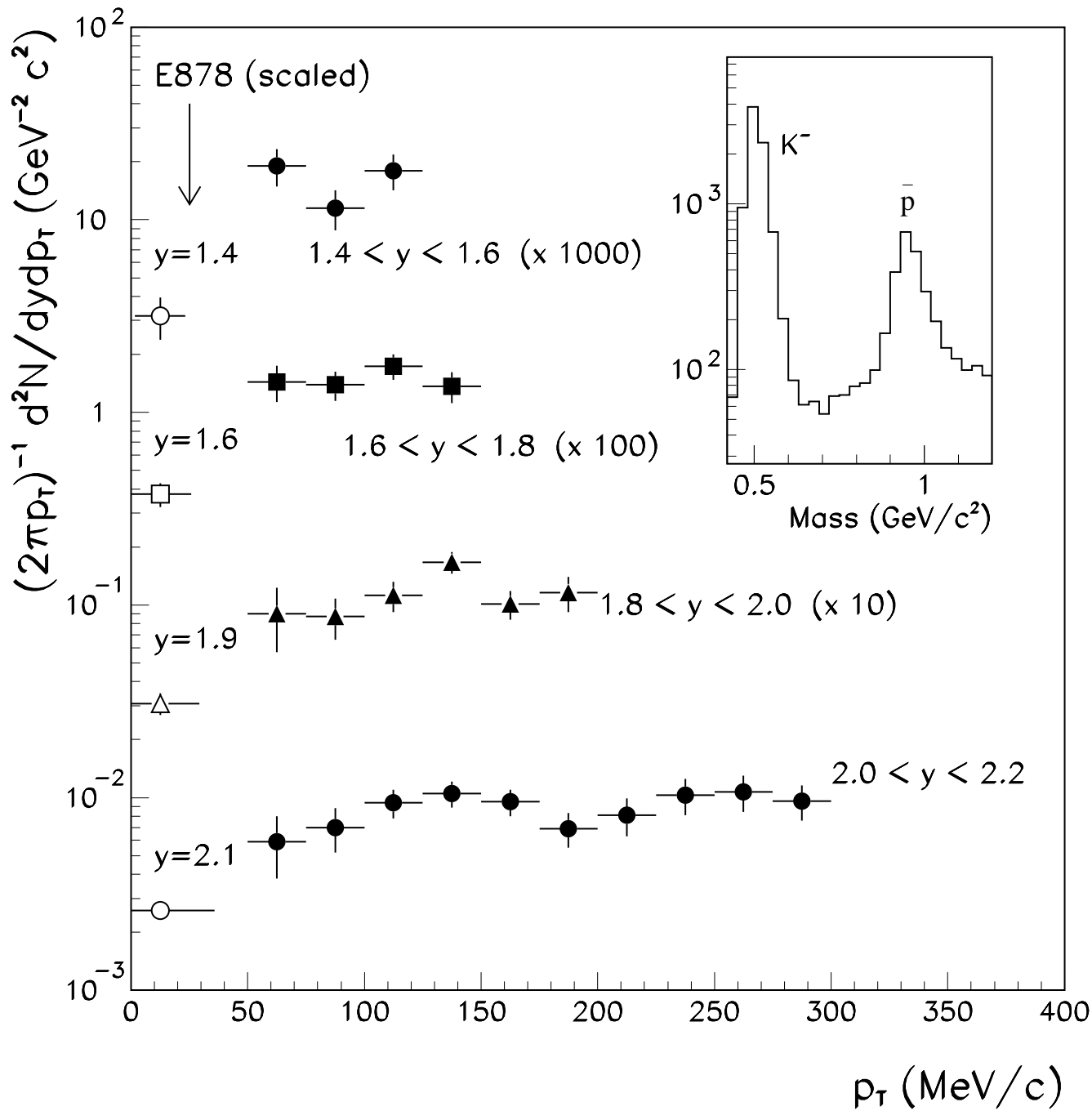
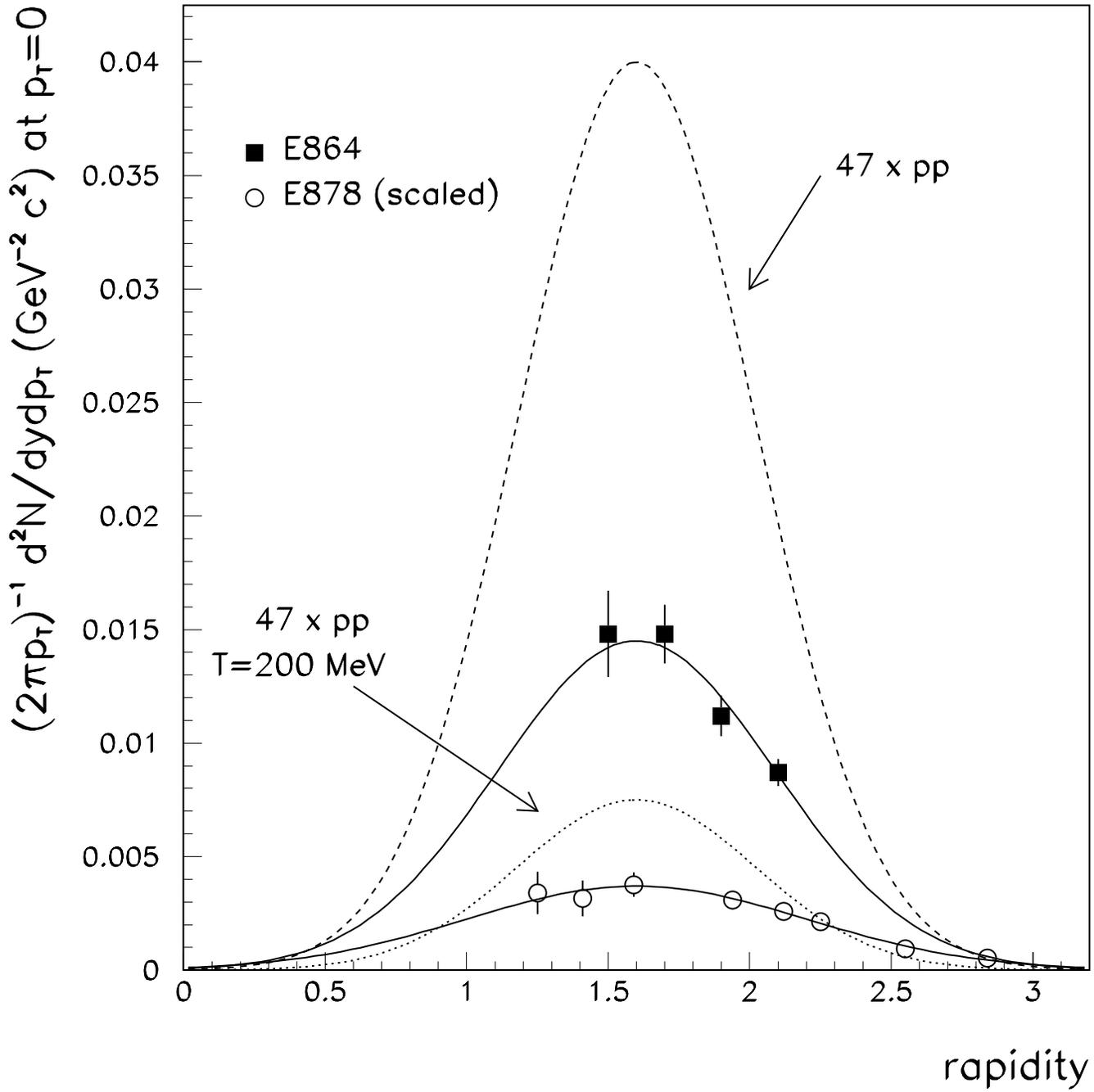
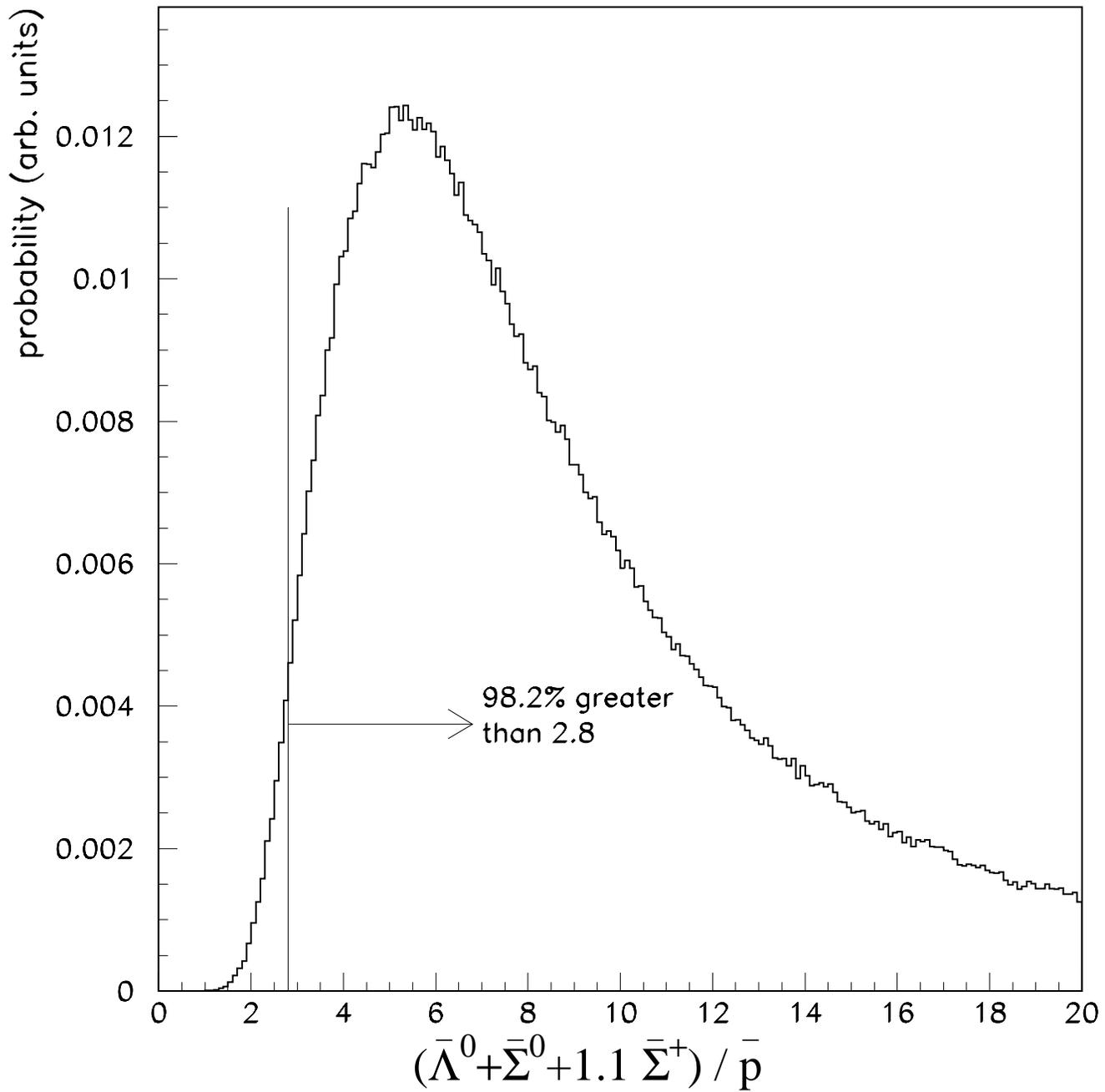
\end{document}